\documentclass[twocolumn,superscriptaddress,showpacs,amsmath,amssymb,pra,nofootinbib]{revtex4-1}    %Two column layout

\usepackage{graphicx} % Include figure files
\usepackage{dcolumn}  % Align table columns on decimal point
\usepackage{bm}       % bold math
\usepackage[usenames,dvipsnames]{color}
\definecolor{URLCOL}{rgb}{0,0.3,0.7} %external link color
\definecolor{LINKCOL}{rgb}{0.05,0.5,0} %internal link color
\definecolor{CITECOL}{rgb}{0.25,0,0.48} %link to bibliography
\usepackage[bookmarks,breaklinks,bookmarksopen,bookmarksnumbered,colorlinks,linkcolor=LINKCOL,linktocpage,citecolor=CITECOL,urlcolor=URLCOL,pdfpagemode=UseOutline,pdftex,pagebackref]{hyperref}

\usepackage{fancyhdr}
\usepackage{booktabs}  % nice tables
\usepackage{dcolumn}
\newcolumntype{d}{D{.}{.}{-1}}
\usepackage{natbib}
\newcommand{\br}{{\bm{r}}}

\def\tocsecspace#1{}
\def\tocless{}

\definecolor{TITLECOL}{rgb}{0.1,0.2,0.7} %title color
\definecolor{CHAPCOL}{rgb}{0,0.48,0} %chapter color
\definecolor{SECOL}{rgb}{0.1,0.2,0.7} %sec color
\definecolor{SHDCOL}{rgb}{0.4,0,0} % heading of section color

\def\sec#1{\tocsecspace{20pt}\section{\textcolor{SECOL}{#1}}
\markboth{\textcolor{SECOL}{\em \bf #1}}{}
}
\def\ssec#1{\tocsecspace{10pt}\tocless\subsection{\textcolor{SSECOL}{#1}}
\markright{\textcolor{SSECOL}{#1}}
}

\def\coloredtitle#1{\title{\textcolor{TITLECOL}{#1}}} %title color
\def\coloredauthor#1{\author{\textcolor{CITECOL}{#1}}} %author color
\def\coltableofcontents{ %table of contents with different colors
\definecolor{SECOL}{rgb}{0.25,0,0.48} %define secol as the subsection color
\definecolor{SSECOL}{rgb}{0.2,0.08,0.53} %subsubsection color  0.2,0.08,0.53
\tableofcontents
\definecolor{SECOL}{rgb}{0.1,0.2,0.7} %put sec color back to what it was
\definecolor{SSECOL}{rgb}{0.25,0,0.48} %ssection color
}

\def\n{n}

\def\mnote#1{\marginpar{\footnotesize\raggedright \textcolor{red}{#1}}}

\definecolor{ORGCOL}{rgb}{0.8,0.65,0.1} %sec color

\def\bea{\begin{eqnarray}}
\def\eea{\end{eqnarray}}
\def\ben{\begin{equation}}
\def\een{\end{equation}}
\def\benu{\begin{enumerate}}
\def\enu{\end{enumerate}}

\def\bei{\begin{itemize}}
\def\eei{\end{itemize}}
\def\beit{\begin{itemize}}
\def\eit{\end{itemize}}
\def\benu{\begin{enumerate}}
\def\enu{\end{enumerate}}

\def\mnote#1{}

\newcommand{\Id}[1] {\int \! \! {\rm d}^3 #1}
\newcommand{\bra}[1] { \langle #1 | }
\newcommand{\ket}[1] { | #1 \rangle }
\newcommand{\braket}[1] { \langle #1 |  #1 \rangle}

\def\F{F}

\def\n{n}

\begin{document}
\coloredtitle{Thermal Density Functional Theory in Context}
\thanks{\textcolor{SHDCOL}{\footnotesize {\bf Appearing in: }\\
{{\em Computational Challenges in Warm Dense Matter}}, \\
edited by F. Graziani, {\em et al.} (Springer, to be published).
}}
%\titlerunning{\Large The Role of Exact Conditions in TDDFT}
\coloredauthor{Aurora Pribram-Jones}
\email{apribram@uci.edu}
\affiliation{Department of Chemistry, University of California, Irvine, CA 92697, USA}
\coloredauthor{Stefano Pittalis}
\affiliation{Department of Chemistry, University of California, Irvine, CA 92697, USA and CNR--Istituto Nanoscienze, Centro S3, Via Campi 213a, I-41125, Modena, Italy}
\coloredauthor{E.K.U. Gross}
\affiliation{Max-Planck-Institut f\"{u}r Mikrostrukturphysik, Weinberg 2, D-06120 Halle, Germany}
\coloredauthor{Kieron Burke}
\affiliation{Departments of Physics and Chemistry, University of California, Irvine, CA 92697, USA}
%\authorrunning{L.~O.~Wagner and K.~Burke}
%\tocauthor{L.~O.~Wagner and K.~Burke}

%\date{\today}
%

\pagestyle{fancy}
\maketitle              % typesets the title of the contribution

\coltableofcontents

\sec{Abstract}
This chapter introduces thermal density functional theory, starting from the ground-state theory and assuming 
a background in quantum mechanics and statistical mechanics.  
We review the foundations of density functional theory (DFT) by illustrating some of its key reformulations. The basics of DFT for thermal ensembles are explained in this context, as are tools useful for analysis and development of approximations.  We close by discussing some key ideas relating thermal DFT and the ground state.  This review emphasizes thermal DFT's strengths as a consistent and general framework.

\sec{Introduction}
\label{sec:1}
The subject matter of high-energy-density physics is vast~\cite{HEDP03}, and the
various methods for modeling it are diverse~\cite{MD06,GBBC12,STVM00}.  The field includes enormous temperature, pressure, and density ranges, reaching regimes where the tools of plasma physics are appropriate~\cite{A04}.  But, especially nowadays, interest also stretches down to warm dense matter (WDM), where chemical details
can become not just relevant, but vital~\cite{KD09}.  WDM, in
turn, is sufficiently close to zero-temperature, ground-state electronic structure that the methods from that field, especially
Kohn-Sham density functional theory (KS DFT)~\cite{KRDM08,RMCH10}, provide a standard
paradigm for calculating material-specific properties with useful accuracy.

It is important to understand, from the outset, that the logic
and methodology of KS-DFT is at times foreign to other
techniques of theoretical physics.  The procedures of KS-DFT
appear simple, yet the underlying theory is surprisingly
subtle.  Consequently, progress in developing useful approximations,
or even writing down formally correct expressions, has been
incredibly slow.  As the KS methodology develops in WDM
and beyond, it is worth taking a few moments to wrap one's
head around its logic, as it does lead to one of the
most successful paradigms of modern electronic structure
theory~\cite{B12}.

This chapter sketches how the methodology
of KS DFT can be generalized to warm systems, and what new features
are introduced in doing so.  It is primarily designed for those
unfamiliar with DFT to get a general understanding of how it functions
and what promises it holds in the domain of warm dense
matter. 
Section 2 is a general review of the basic theorems of DFT, using the
original methodology of Hohenberg-Kohn~\cite{HK64} and then the more general Levy-Lieb construction~\cite{L79,L83}.   In Section 3, we discuss approximations, which are always necessary
in practice, and several important exact conditions that are used
to guide their construction.   In Section 4, we review the thermal KS equations~\cite{M65} and some relevant statistical mechanics.  Section 5 summarizes some of the most important exact conditions for thermal ensembles~\cite{PPFS11,DT11}.  Last, but not least, in Section 6
we review some recent results that generalize ground-state exact scaling
conditions and note some of the main differences between the finite-temperature and the ground-state formulation.

\sec{Density functional theory}
\label{sec:2}

A reformulation of the interacting many-electron problem in terms of the electron density rather than the many-electron wavefunction has been attempted since the early days of quantum mechanics~\cite{T27,F27,F28}. The advantage is clear: while the wavefunction for interacting electrons depends in a complex fashion on all the particle coordinates, the particle density is a function of only three spatial coordinates.

Initially, it was believed that formulating quantum mechanics solely in terms of the particle density gives only an approximate solution, as in the Thomas-Fermi method~\cite{T27, F27,F28}. However, in the mid-1960s, Hohenberg and Kohn~\cite{HK64} showed that, for systems of electrons in an external potential, all the properties of the many-electron ground state are, in principle, exactly determined by the ground-state particle density alone.

Another important approach to the many-particle problem appeared early in the development of quantum mechanics: the single-particle approximation. Here, the two-particle potential representing the interaction between particles is replaced by some effective, one-particle potential. A prominent example of this approach is the Hartree-Fock method~\cite{F30,H35}, which includes only exchange contributions in its effective one-particle potential. A year after the Hohenberg-Kohn theorem had been proven, Kohn and Sham~\cite{KS65} took a giant leap forward. They took the ground state particle density as the basic quantity and showed that
both exchange and correlation effects due to the electron-electron interaction can be treated through an effective single-particle Schr\"odinger equation. Although Kohn and Sham wrote their paper using the local density approximation, they also pointed out the exactness of that scheme if the exact exchange-correlation functional were to be used (see Section \ref{subsec:3}).  The KS scheme is used in almost all DFT calculations of electronic structure today.  Much development in this field remains focused on improving approximations to the exchange-correlation energy (see Section \ref{sec:3}).

The Hohenberg-Kohn theorem and Kohn-Sham scheme are the basic elements of modern density-functional theory (DFT)~\cite{B12, B07, BW13}. We will review the initial formulation of DFT for non-degenerate ground states and its later extension to degenerate ground states. Alternative and refined mathematical formulations are then introduced.

\ssec{Introduction}
\label{subsec:1}

The non-relativistic Hamiltonian\footnote{See Refs.~\cite{Schwabl07} or ~\cite{Sakurai93} for quantum mechanical background that is useful for this chapter.} for $N$ interacting electrons\footnote{In this work, we discuss only spin-unpolarized electrons.} moving in a static potential $v({\bf r})$ reads (in atomic units)
\begin{equation}
\hat{H} = \hat{T} + \hat{V}_{ee} + \hat{V} := - \frac{1}{2} \sum_{i=1}^N \nabla^2_i + 
\frac{1}{2} \sum_{\stackrel{i,j=1}{i \neq j}}^N 
\frac{1}{| {\bf r}_i - {\bf r}_j|} + \sum_{i=1}^N v({\bf r}_i).
\end{equation}
Here, $\hat{T}$ is the total kinetic-energy operator, $\hat{V}_{ee}$ describes the repulsion between the electrons, and $\hat{V}$ is a local (multiplicative) scalar operator. This includes the interaction of the electrons with the nuclei 
(considered within the Born-Oppenheimer approximation) and any other external scalar potentials.

The eigenstates, $\Psi_{i}({\bf r}_1,...,{\bf r}_N)$, of the system are obtained by solving the eigenvalue problem
\begin{equation}
\label{EGVP}
\hat{H} \Psi_{i}({\bf r}_1,...,{\bf r}_N) = E_i \Psi_{i}({\bf r}_1,...,{\bf r}_N),
\end{equation}
with appropriate boundary conditions for the physical problem at hand.
Eq. (\ref{EGVP}) is the time-independent Schr\"odinger equation.
We are particularly interested in the ground state, the eigenstate
with lowest energy, and assume the wavefunction can be normalized.

Due to the interactions among the electrons, $\hat{V}_{ee}$, an explicit and closed solution of the many-electron problem in Eq. (\ref{EGVP}) is, in general, not possible. But because accurate prediction of a wide range of physical and chemical phenomena requires inclusion of electron-electron interaction, we need a path to accurate approximate solutions.

Once the number of electrons with Coulombic interaction is given, the Hamiltonian is determined by specifying the external potential. For a given $v({\bf r})$, the total energy is a functional of the many-body wavefunction $\Psi({\bf r}_1,...,{\bf r}_N)$
\begin{equation}
\label{MBE2}
E_v[\Psi] = \bra{\Psi} \hat{T} + \hat{V}_{ee} + \hat{V} \ket{\Psi}\;.
\end{equation}
The energy functional in Eq. (\ref{MBE2}) may be evaluated for any $N$-electron wavefunction, and the Rayleigh-Ritz variational principle ensures that 
the ground state energy, $E_{v}$, is given by
\begin{equation}
\label{RR}
E_{v} = \inf_{\Psi} E_{v}[\Psi],
\end{equation}
where the infimum is taken over all normalized, antisymmetric wavefunctions. The Euler-Lagrange equation expressing the minimization of the energy is
\begin{equation}
\label{NCRR}
\frac{\delta}{\delta \Psi} \left\{ 
E_{v}[\Psi] - 
\mu \left[ \braket{\Psi} - 1 \right]\right\} = 0,
\end{equation}
where the functional derivative is performed over $\Psi \in {\cal L}^2(\mathbb{R}^{3N})$ (defined as in Ref.~\cite{ED11}). Relation (\ref{NCRR}) again leads to the many-body Schr\"odinger equation and the Lagrangian multiplier $\mu$ can be identified as the chemical potential. 

We now have a procedure for finding approximate solutions by restricting the
form of the wavefunctions. In the Hartree-Fock (HF) approximation, for example, the form of the wave-function is restricted to a single
Slater determinant. Building on the HF wavefunction, modern quantum chemical methods can produce extremely accurate solutions to the Schr\"{o}dinger equation~\cite{S10}.  Unfortunately, wavefunction-based approaches that go beyond HF usually are afflicted by an impractical growth of the numerical effort with the number of particles. Inspired by the Thomas-Fermi approach, one might wonder if the role played by the wavefunction could be played by the particle density, defined as 
\begin{align}
\label{ND}
n({\bf r}) &:=  \bra{\Psi} \sum_{i=1}^{N} \delta({\bf \hat{r}}-{\bf \hat{r}}_{i}) \ket{\Psi}\notag\\
&= N \int d {\bf r}_2 ... \int d {\bf r}_N
\Big| \Psi({\bf r},{\bf r}_2,...,{\bf r}_N)\Big|^2,
\end{align}
from which
\begin{equation}
\label{NDN}
\int {d^3 r} ~n({\bf r})= N.
\end{equation}
In that case, one would deal with a function of only three spatial coordinates, regardless of the number of electrons.

\ssec{Hohenberg-Kohn theorem}
\label{subsec:2}
Happily, the two-part Hohenberg-Kohn (HK) Theorem assures us that the electronic density alone is enough to determine all observable quantities of the systems.  These proofs cleverly connect specific sets of densities, wavefunctions, and potentials, exposing a new framework for the interacting many-body problem.
 
Let $\bf{P}$ be the set of external potentials leading to a {\em non-degenerate} ground state for $N$ electrons. For a given potential, the corresponding ground state, $\Psi$, is obtained through the solution of the Schr\"odinger equation:
\begin{equation}
v ~~ \longrightarrow ~~ \Psi, ~~ \mbox{with} ~~ v \in \bf{P}.
\end{equation}
Wavefunctions obtained this way are called interacting v-representable. 
We collect these ground state wavefunctions in the set $\bf{W}$.
The corresponding particle densities can be computed
using definition (\ref{ND}):
\begin{equation}
\Psi ~~ \longrightarrow ~~ n, ~~ \mbox{with} ~~ \Psi \in \bf{W}.
\end{equation}
Ground state particle densities obtained this way
are also called interacting v-representable. We denote the set of these densities as $\bf{D}$.

\subsubsection{First part}
\begin{figure}[htbp]
\includegraphics[width=\columnwidth]{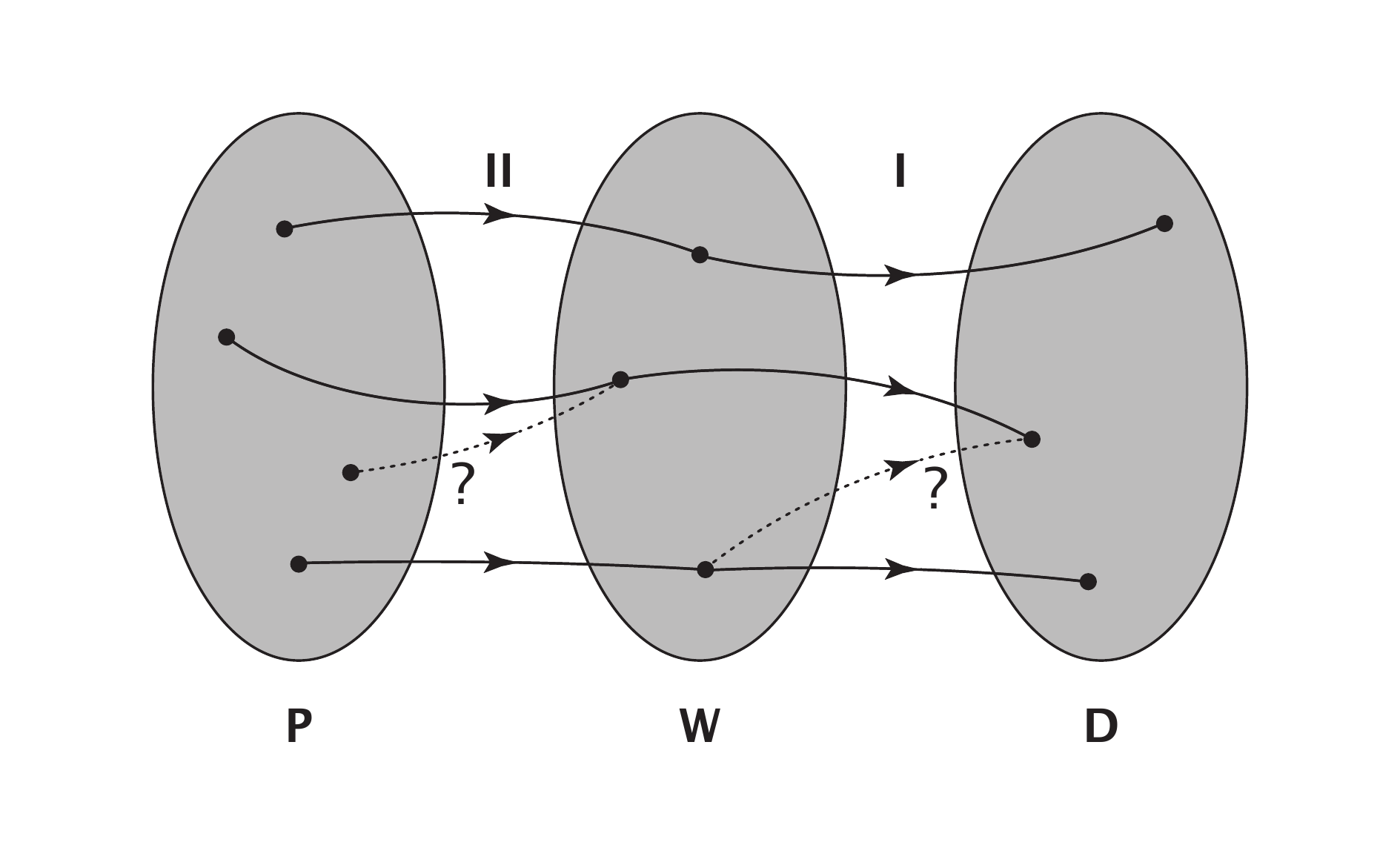}
\caption{The Hohenberg-Kohn proves the one-to-one mappings between potentials and ground-state wavefunctions and between ground-state wavefunctions and ground-state densities.  The dotted lines indicated by question marks show the two-to-one mappings disproved by Hohenberg and Kohn~\cite{ED11,DG90}.
\label{fig:HK}
}
\end{figure}

Given a density $n \in \bf{D}$, the first part of the Hohenberg-Kohn theorem states that the wavefunction $\Psi \in \bf{W}$ leading to $n$ is unique, apart from a constant phase factor. The proof is carried out by {\em reductio ad absurdum} and is illustrated in Figure \ref{fig:HK}.

Consider two different wavefunctions in $\bf{W}$, $\Psi_1$ and $\Psi_2$, that differ by more than a constant phase factor. Next, let $n_1$ and $n_2$ be the corresponding densities computed by Eq. (\ref{ND}). Since, by construction, we are restricting ourselves to non-degenerate ground states, $\Psi_1$ and $\Psi_2$  must come from two different
potentials.  Name these $v_1$ and $v_2$, respectively. 

Assume that these different wavefunctions yield the same density:
\begin{equation}
\label{H210}
\Psi_1 \neq \Psi_2 ~~ \mbox{but}~~ n_1({\bf r}) = n_2({\bf r}).
\end{equation}
Application of the Rayleigh-Ritz variational principle yields the inequality
\begin{equation}
\label{eneq}
\bra{\Psi_1}\hat{H}_1\ket{\Psi_1} < \bra{\Psi_2}\hat{H}_1\ket{\Psi_2}, 
\end{equation}
from which we obtain
\begin{equation}
\label{ineq1}
E_1 < \bra{\Psi_2}\hat{H}_2+(\hat{V}_1-\hat{V}_2)\ket{\Psi_2} =
E_{2} + \int {d^3 r}~n_1({\bf r})\left[  v_1({\bf r}) - v_2({\bf r})  \right].
\end{equation} 
Reversing the role of systems 1 and 2 in the derivation, we find
\begin{equation}
\label{ineq2}
E_2 < \bra{\Psi_1}\hat{H}_1+(\hat{V}_2-\hat{V}_1)\ket{\Psi_1} =
E_{1} + \int {d^3 r}~n_2({\bf r})\left[  v_2({\bf r}) - v_1({\bf r})  \right].
\end{equation}
The assumption that the two densities are equal, $n_1({\bf r}) = n_2({\bf r})$, and addition of the inequalities (\ref{ineq1}) and (\ref{ineq2}) yields
\begin{equation}
E_1 + E_2 < E_1 + E_2,
\end{equation}
which is a contradiction. We conclude that the foregoing hypothesis (\ref{H210}) was wrong, so $n_1 \neq n_2$. Thus each density is the ground-state density of, at most, one wavefunction. This mapping between the density and wavefunction is written
\begin{equation}
\label{HK1}
n ~\longrightarrow~ \Psi,~ \mbox{with} ~n \in {\bf D} ~\mbox{and}~  \Psi \in \bf{W}.
\end{equation}

\subsubsection{Second part}

Having specified the correspondence between density and wavefunction, Hohenberg and Kohn then consider the potential.  By explicitly inverting the Schr\"odinger equation, 
\begin{equation}
\sum_{i=1}^N v({\bf r}_i) = E - \frac{\left( \hat{T} +\hat{V}_{ee} \right)\Psi({\bf r}_1,{\bf r}_2,...,{\bf r}_N)}
{\Psi({\bf r}_1,{\bf r}_2,...,{\bf r}_N)},
\end{equation}
they show the elements $\Psi$ of $\bf{W}$ also determine the elements $v$ of $\bf{P}$, apart from an additive constant.

We summarize this second result by writing
\begin{equation}
\label{HK2}
\Psi ~\longrightarrow~ v,~ \mbox{with} ~\Psi \in {\bf W} ~\mbox{and}~ v \in \bf{P}.
\end{equation}

\subsubsection{Consequences}
\label{cons}

Together, the first and second parts of the theorem yield
\begin{equation}
n \longrightarrow v + const,~~\mbox{with} ~n \in {\bf D} ~\mbox{and}~  v \in \bf{P},
\end{equation}
that the ground state particle density determines the external 
potential up to a trivial additive constant.  This is the first HK theorem.

Moreover, from the first part of the theorem it follows that any ground-state observable is a 
functional of the ground-state particle density. Using the one-to-one dependence 
of the wavefunction, $\Psi[n]$, on the particle density,
\begin{equation}
\bra{\Psi}\hat{O}\ket{\Psi} = \bra{\Psi[n]}\hat{O}\ket{\Psi[n]} = O[n].
\end{equation}
For example, the following functional can be defined:
\begin{align}
\label{HKE}
E_{v,\rm HK}[n] &:= \bra{\Psi[n]}\hat{T}+\hat{V}_{ee}+\hat{V}\ket{\Psi[n]}\notag\\
&=F_{\rm HK}[n] + \int {d^3 r} ~ n({\bf r})v({\bf r}),
\end{align}
where $v$ is a given external potential and $n$ can be any density in $\bf{D}$. Note that
\begin{equation}
\label{FHK}
F_{\rm HK}[n] := \bra{\Psi[n]} \hat{T} + \hat{V}_{ee} \ket{\Psi[n]}
\end{equation}
is independent of $v$. The second HK theorem is simply that $F_{\rm HK}[n]$ is independent of $v({\bf r})$. This is therefore a universal functional of the ground-state particle density.  We use the subscript, ${\rm HK}$, to emphasize that this is the original density functional of Hohenberg and Kohn.

Let $n_0$ be the ground-state particle density of the potential $v_0$. The Rayleigh-Ritz variational principle (\ref{RR}) immediately tells us
\begin{equation}
\label{TDVP}
E_{v_0} = \min_{n \in \bf{D}} E_{v_0,\rm HK}[n] = E_{v_0,\rm HK}[n_0].
\end{equation}
We have finally obtained a variational principle based on the particle density instead of the computationally expensive wavefunction.

\subsubsection{Extension to degenerate ground states}

The Hohenberg-Kohn theorem can be generalized by allowing $\bf{P}$ to include local potentials having {\em degenerate} ground states~\cite{L79,Kohn:85,DG90}, . This means an entire subspace of wavefunctions can correspond to the lowest eigenvalue of the Schr\"odinger equation (\ref{EGVP}). The sets $\bf{W}$ and $\bf{D}$ are enlarged accordingly, to include all the additional ground-state wavefunctions and particle densities.

\begin{figure}[htbp]
\includegraphics[width=\columnwidth]{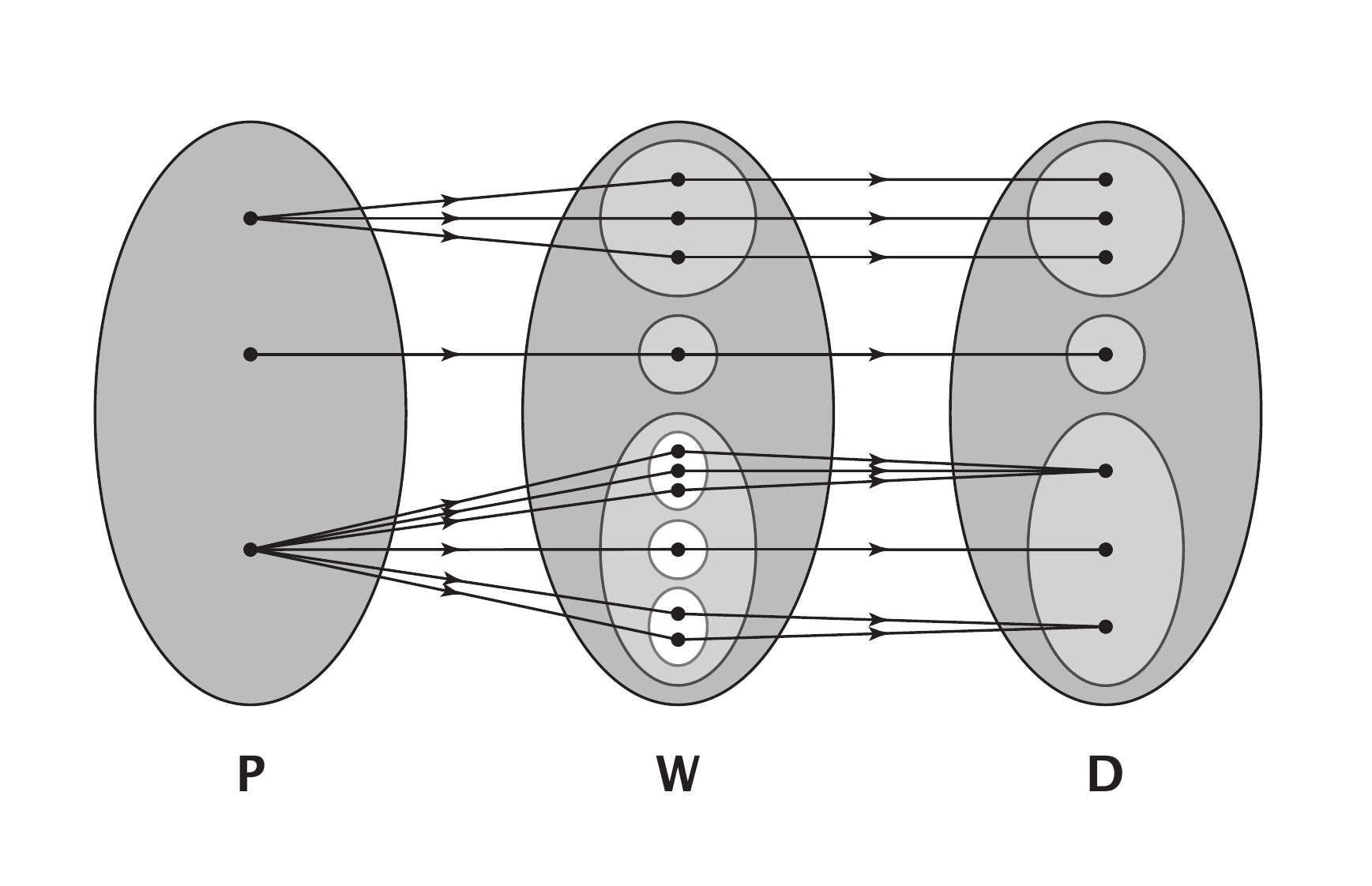}
\caption{The mappings between sets of potentials, wavefunctions, and densities can be extended to include potentials with degenerate ground states.  This is seen in the one-to-many mappings between $\bf{P}$ and $\bf{W}$.  Note also the many-to-one mappings from $\bf{W}$ to $\bf{D}$ caused by this degeneracy~\cite{GV08,DG90}.
\label{fig:degcase}
}
\end{figure}

In contrast to the non-degenerate case, the solution of the Schr\"odinger equation (\ref{EGVP}) now establishes a mapping from $\bf{P}$ to $\bf{W}$ which is one-to-many (see Figure \ref{fig:degcase}). Moreover, different degenerate wavefunctions can have the same particle density. Equation (\ref{ND}), therefore, establishes a mapping from $\bf{W}$ to $\bf{D}$ that is many-to-one. However, any one of the degenerate ground-state densities still determines the potential uniquely.

The first part of the HK theorem needs to be modified in light of this alteration of the mapping between wavefunctions and densities. To begin, note that two degenerate subspaces, sets of ground states of two different potentials, are disjoint. Assuming that a common eigenstate $\Psi$ can be found, subtraction of one Schr\"odinger equation from the other yields 
\begin{equation}
(\hat{V}_1-\hat{V}_2) \Psi = (E_1 - E_2) \Psi.
\end{equation}
For this identity to be true, the eigenstate $\Psi$ must vanish in the region where
the two potentials differ by more than an additive constant. This region has measure greater than zero. Eigenfunctions of potentials in $\bf{P}$, however, vanish only on sets of measure zero~\cite{L85}. This contradiction lets us conclude that $v_1$ and $v_2$ cannot have common eigenstates. We then show that ground states from two different potentials always have different particle densities using the Rayleigh-Ritz variational principle as in the non-degenerate case. 

However, two or more degenerate ground state wavefunctions can have the same particle density. As a consequence, neither the wavefunctions nor a generic ground state property can be determined uniquely from knowledge of the ground state particle density alone. This demands reconsideration of the definition of the universal $F_{\rm HK}$ as well. Below, we verify that the definition of $F_{\rm HK}$ does not rely upon one-to-one correspondence among ground state wavefunctions and particle densities.

The second part of the HK theorem in this case proceeds as in the original proof, with each ground state in a degenerate level determining the external potential up to an additive constant. Combining the first and second parts of the proof again confirms that any element of $\bf{D}$ determines an element of $\bf{P}$, up to an additive constant. In particular, any one of the degenerate densities determines the external potential. Using this fact and that the total energy is the same for all wavefunctions in a given degenerate level, we define $F_{\rm HK}$:
\begin{equation}
\label{HKE2}
F_{\rm HK}[n] :=  E\left[v[n]\right] - \int {d^3 r}~v[n]({\bf r}) n({\bf r}).
\end{equation}
This implies that the value of
\begin{equation}
\label{FHK2}
F_{\rm HK}[n] = \bra{\Psi_0 \rightarrow n} \hat{T} + \hat{V}_{ee} \ket{\Psi_0 \rightarrow n}
\end{equation}
is the same for all degenerate ground-state wavefunctions that
have the same particle density. The variational principle based on the particle density can then be formulated as before in Eq. (\ref{TDVP}).

\ssec{Kohn-Sham scheme}
\label{subsec:3}

The exact expressions defining $F_{\rm HK}$ in the previous section
 are only formal ones. In practice, $F_{\rm HK}$ must be approximated. Finding approximations that yield usefully accurate results 
turns out to be an extremely difficult task, so much so that pure, orbital-free approximations for $F_{\rm HK}$ are not pursued in most modern DFT calculations. Instead, efficient approximations can be constructed by introducing the Kohn-Sham scheme, in which a useful decomposition of $F_{\rm HK}$ in terms of other density functionals is introduced. In fact, the Kohn-Sham decomposition is so effective that effort on orbital-free DFT utilizes the Kohn-Sham structure, but not its explicitly orbital-dependent expressions.

Consider the Hamiltonian of $N$ non-interacting electrons
\begin{equation}
\label{HNI}
\hat{H}_s = \hat{T} + \hat{V} := - \frac{1}{2} \sum_{i=1}^N \nabla^2_i
+ \sum_{i=1}^N v({\bf r}_i).
\end{equation}
Mimicking our procedure with the interacting system, we group external local potentials in the set $\bf{P}$. The corresponding non-interacting ground state wavefunctions $\Psi_s$ are then grouped in the set $\bf{W^s}$, and their particle densities $n_s$ are grouped in $\bf{D^s}$. We can then apply the HK theorem and define the non-interacting analog of $F_{\rm HK}$, which is simply the kinetic energy:
\begin{equation}
\label{TSG}
T_{s}[n_s] :=  E\left[v[n_s]\right] - \int {d^3 r} ~v[n_s]({\bf r}) n_s({\bf r}).
\end{equation}
Restricting ourselves to non-degenerate ground states, the expression in Eq. (\ref{TSG}) can be rewritten to stress the one-to-one correspondence among densities and wavefunctions:
\begin{equation}
\label{TSG2}
T_{s}[n_s] = \bra{\Psi_s[n_s]} \hat{T} \ket{\Psi_s[n_s]}\;.
\end{equation}
We now introduce a fundamental assumption: for each element $n$ of $\bf{D}$, a potential $v_{s}$ in $\bf{P^s}$ exists, with corresponding ground-state particle density $n_{s}=n$. We call $v_s$ the Kohn-Sham potential. In other words, interacting v-representable densities are also assumed to be non-interacting v-representable. This maps the interacting problem onto a non-interacting one.

Assuming the existence of $v_{s}$, the HK theorem
applied to the class of non-interacting systems ensures that $v_{s}$ is
unique up to an additive constant. As a result, we find the particle density of the interacting system by solving the non-interacting eigenvalue problem, which is called the Kohn-Sham equation:
\begin{equation}
\label{KSG}
\hat{H}_s \Phi = E\Phi.
\end{equation}
For non-degenerate ground states, the Kohn-Sham ground-state wavefunction is a single Slater determinant. In general, when considering degenerate ground states, the Kohn-Sham wavefunction can be expressed as a linear combination of several Slater determinants~\cite{L83,EE83}. There also exist interacting ground states with particle densities that can only be represented by an ensemble of non-interacting particle densities~\cite{Averill:92,Wang:96,Schipper:98,Schipper:99,UK01}. We will
come back to this point in Section \ref{ensform}.

Here we continue by considering the simplest cases of non-degenerate ground states. Eq. (\ref{KSG}) can be rewritten in terms of the single-particle orbitals as follows:
\begin{equation}
\label{KS1}
\left[-\frac{1}{2}
\nabla^2+v_s({\bf r})\right] \varphi_{i}({\bf r})= \epsilon_{i} \varphi_{i}({\bf r})\;.
\end{equation}
The single-particle orbitals $\varphi_{i}({\bf r})$ are called Kohn-Sham orbitals and Kohn-Sham wavefunctions are Slater determinants of these orbitals. Via the Kohn-Sham equations, the orbitals are implicit functionals of $n({\bf r})$. We emphasize that -- although in DFT the particle density is the only basic variable -- the Kohn-Sham orbitals are proper fermionic single-particle states. The ground-state Kohn-Sham wavefunction is obtained by occupying the $N$ eigenstates with lowest eigenvalues. The corresponding density is
\begin{equation}
n({\bf r}) =\sum_{i=1}^{N} n_i |\varphi_{i}({\bf r})|^2,
\end{equation}
with $n_i$ the $i^{\rm th}$ occupation number.

In the next section, we consider the consequences of introducing the
Kohn-Sham system in DFT. 

\subsubsection{Exchange-correlation energy functional}
A large fraction of $F_{\rm HK}[n]$ can be expressed in terms of kinetic and electrostatic energy. This decomposition
is given by
\begin{equation}
\label{KSDECF}
F_{\rm HK}[n] = T_{s}[n] + U[n] + E_{xc}[n]~.
\end{equation}
The first term is the kinetic energy of the Kohn-Sham system,
\begin{equation}\label{TSG3}
T_{s}[n] = - \frac{1}{2}\sum_{i=1}^{N}  \Id{r} \; \varphi^*_{i}({\bf r}) \nabla^2
\varphi_{i}({\bf r})\;.
\end{equation}
The second is the Hartree energy (a.k.a. electrostatic self-energy, a.k.a. Coulomb energy), 
\begin{equation}
\label{UH}
U[n] = \frac{1}{2} \int \int {d^3 r} {d^3 r'} ~ \frac{n({\bf r})n({\bf r'})}{|{\bf r}-{\bf r'}|}\;.
\end{equation}
The remainder is defined as the exchange-correlation energy,
\begin{equation}
\label{DEX}
E_{xc}[n] := F_{\rm HK}[n] - T_{s}[n] - U[n]~.
\end{equation}
For systems having more than one particle, $E_{xc}$ accounts for exchange and correlation energy contributions. Comparing Eqs. (\ref{KSDECF}) and (\ref{HKE}), the total energy density functional is
\begin{equation}
\label{DEF}
E_{v,\rm HK}[n] = T_{s}[n] + U[n] + E_{xc}[n] + \int {d^3 r}~n({\bf r})v({\bf r}).
\end{equation}

Consider now the Euler equations for the interacting and non-interacting system.  Assuming the differentiability of the functionals (see Section \ref{ensform}), these necessary conditions for having energy minima are
\begin{equation}
\label{EQ1}
\frac{\delta F_{\rm HK}}{\delta n({\bf r})} + v({\bf r}) = 0 % -\mu = 0,
\end{equation}
and
\begin{equation}
\label{EQ2}
\frac{\delta T_s}{\delta n({\bf r})} + v_s({\bf r}) = 0, % -\mu_s = 0,
\end{equation}
respectively. With definition (\ref{KSDECF}), from Eqs. (\ref{EQ1}) and (\ref{EQ2}), we obtain
\begin{equation}
\label{VS}
v_s({\bf r}) = v_{H}[n]({\bf r}) + v_{xc}[n]({\bf r}) + v({\bf r}).
\end{equation}
Here, $v({\bf r})$ is the external potential acting upon the interacting electrons,
$v_{H}[n]({\bf r})$ is the Hartree potential,
\begin{equation}
\label{VH}
v_{H}[n]({\bf r}) = \int {d^3 r'} \frac{n({\bf r'})}{|{\bf r}-{\bf r'}|}=\frac{\delta U}{\delta n({\bf r})},
\end{equation}
and $v_{xc}[n]({\bf r}) $ is the exchange-correlation potential,
\begin{equation}
\label{VXC}
v_{xc}[n]({\bf r}) = \frac{\delta E_{xc}[n]}{\delta n({\bf r})}.
\end{equation}

Through the decomposition in Eq. (\ref{KSDECF}), a significant part of $F_{\rm HK}$ is in the explicit form of $T_{s}[n] + U[n]$ without approximation. Though often small, the $E_{xc}$ density functional still represents an important part of the total energy.  Its exact functional form is unknown, and it therefore must be approximated in practice. However, good and surprisingly efficient approximations exist for $E_{xc}$.

We next consider reformulations of DFT,  which allow analysis and solution of some important technical questions at the heart of DFT. They also have a long history of influencing the analysis of properties of the exact functionals. 

\ssec{Levy's formulation}
An important consequence of the HK theorem is that the Rayleigh-Ritz variational principle based on the wavefunction can be replaced by a variational principle based on the particle density. The latter is valid for all densities in the set $\bf{D}$, the set of v-representable densities. Unfortunately,  v-representability is a condition which is not easily verified for a given function $n({\bf r})$. Hence it is highly desirable to formulate the variational principle over a set of densities characterized by simpler conditions. This was provided by Levy~\cite{L79} and later reformulated and extended by Lieb~\cite{L83}. In this and the sections that follow, Lebesgue and Sobolev spaces are defined in the usual way~\cite{ED11, RS81}.

First, the set $\bf{W}$ is enlarged to $\bf{W_N}$, which includes all possible antisymmetric and normalized $N$-particle wavefunctions $\Psi$. The set $\bf{W_N}$ now also contains $N$-particle wavefunctions which are {\em not} necessarily ground-state wavefunctions to some external potential $v$, though it remains in the same Sobolev space~\cite{ED11} as $\bf{W}$: ${\cal H}^1(\mathbb{R}^{3N})$. Correspondingly, the set $\bf{D}$ is replaced by the set $\bf{D_N}$. $\bf{D_N}$ contains the densities generated from the $N$-particle antisymmetric wavefunctions in $\bf{W_N}$ using Eq. (\ref{ND}):

\begin{equation}
{\bf D_N}=\left\{n\mid n({\bf r})\geq 0,\int d^3r~n({\bf r})=N, n^{1/2}({\bf r})\in {\cal H}^1(\mathbb{R}^{3})\right\}.
\end{equation}
The densities of $\bf{D_N}$ are therefore called $N$-representable. Harriman's explicit construction~\cite{H81} shows that any integrable and positive function $n({\bf r})$ is $N$-representable.

\begin{figure}[htbp]
\includegraphics[width=\columnwidth]{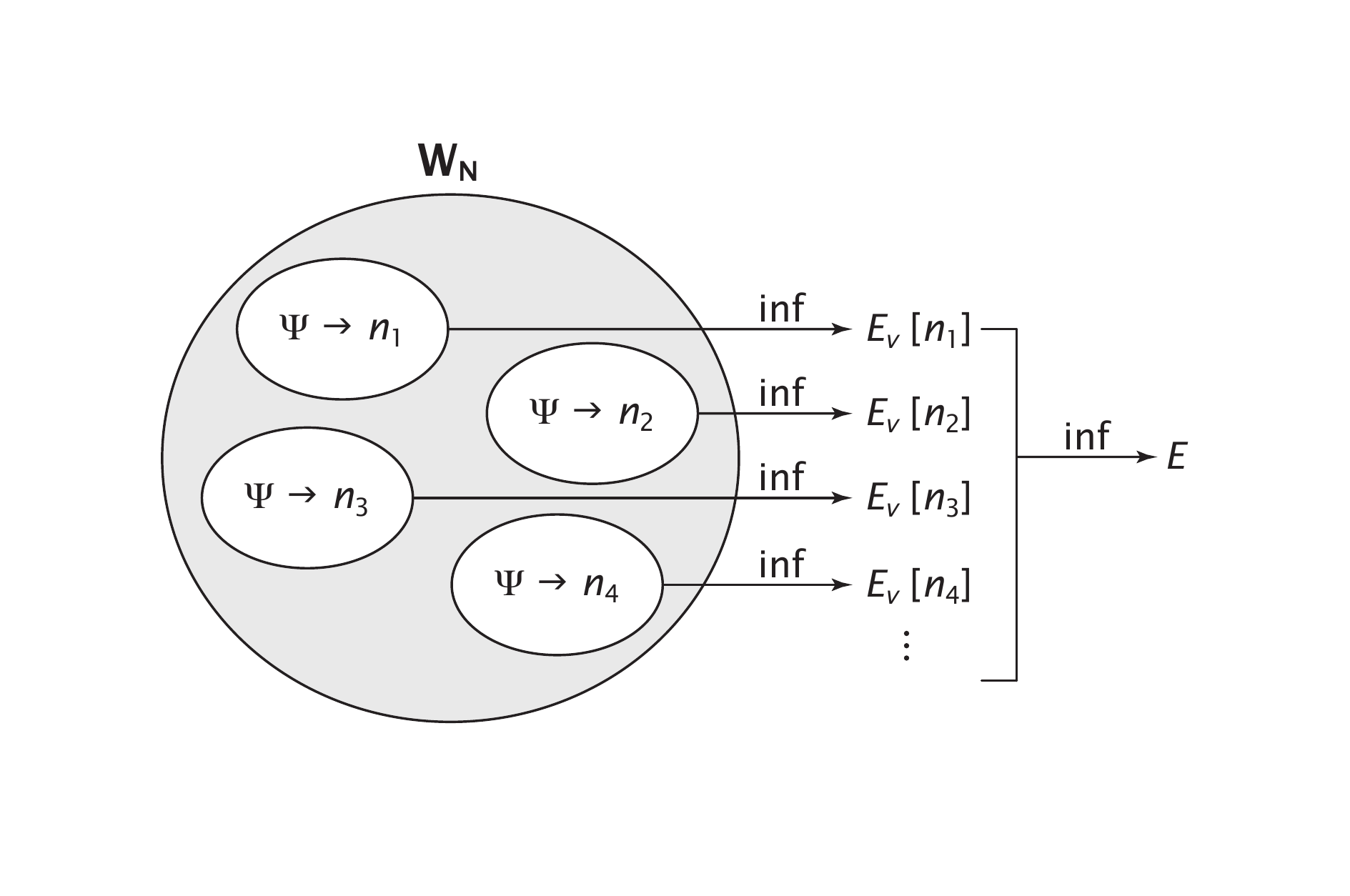}
\caption{This diagram shows the two-step minimization of Levy's constrained search.  The first infimum search is over all wavefunctions corresponding to a certain density $n_i$.  The second search runs over all of the densities~\cite{GV08,PY89}.
\label{fig:consrch}
}
\end{figure}

Levy reformulated the variational principle in a constrained-search fashion (see Figure \ref{fig:consrch}):
\begin{equation}
E_v = \inf_{n \in \bf{D_N}}~~ \left\{  \inf_{\Psi \rightarrow n | \Psi \in \bf{W_N}}  
\bra{\Psi} \hat{T} + \hat{V}_{ee} \ket{\Psi} + \int {d^3 r}~n({\bf r})v({\bf r}) \right\}.
\end{equation}
In this formulation, the search inside the braces is constrained to those wavefunctions which yield a given density $n$  -- therefore the name ``constrained search". The minimum is then found by the outer search over all densities. The potential $v({\bf r})$ acts like a Lagrangian multiplier to satisfy the constraint on the density at each point in space. In this formulation, $F_{\rm HK}$  is replaced by
\begin{equation}
\label{LL}
F_{\rm LL}[n] := \inf_{\Psi \rightarrow n} \bra{\Psi} \hat{T} + \hat{V}_{ee} \ket{\Psi},~~ \mbox{with}~ 
\Psi \in {\bf{W_N}} ~\mbox{and}~ n \in \bf{D_N}\;.
\end{equation}
The functional $E_{\rm HK}$ can then be replaced by
\begin{equation}
E_{v,\rm LL}[n] := F_{\rm LL}[n] + \int {d^3 r}~n({\bf r})v({\bf r}), ~~ \mbox{with}~ n \in \bf{D_N}.
\end{equation}
If, for a given $v_0$, the corresponding ground-state
particle density, $n_0$, is inserted, then
\begin{equation}
E_{v_0,\rm LL}[n_0] = E_{v_0,\rm HK}[n_0] = E_{v_0},
\end{equation}
from which  
\begin{equation}
F_{\rm LL}[n] = F_{\rm HK}[n],~~ \mbox{for~all}~n~\in~\bf{D}~.
\end{equation}
Furthermore, if any other particle density is inserted, we obtain
\begin{equation}
E_{v_0,\rm LL}[n] \ge E_{v_0},~~ \mbox{for}~~ n \ne n_0~~\mbox{and}~ n \in \bf{D_N}.
\end{equation}

In this approach, the degenerate case does not require particular care. 
In fact, the correspondences between potentials, wavefunctions and densities are not explicitly employed as they were in the previous Hohenberg-Kohn formulation. However, the $N$-representability is of secondary importance in the context of the Kohn-Sham scheme.  There, it is still necessary to assume that the densities of the interacting electrons are non-interacting v-representable as well. We discuss this point in more detail in the next section.

Though it can be shown that the $F_{\rm LL}[n]$ infimum is a minimum~\cite{L83}, the functional's lack of convexity causes a serious problem in proving the differentiability of $F_{\rm LL}$~\cite{L83}.  Differentiability is needed to define an Euler equation for finding $n({\bf r})$ self-consistently. This is somewhat alleviated by the Lieb formulation of DFT (see below). 

\ssec{Ensemble-DFT and Lieb's formulation}
\label{ensform}
In the remainder of this section, 
we are summarizing more extensive and pedagogical reviews that can be found in Refs.~\cite{ED11},~\cite{DG90}, and~\cite{L03}. Differentiability of functionals is, essentially, related to the convexity of the functionals. 
Levy and Lieb showed that the set $\bf{D}$ is not convex~\cite{L83}.  In fact, there exist combinations of the form
\begin{equation}
\label{ED}
n({\bf r}) =\sum_{k=1}^{M} \lambda_k n_k({\bf r}),~ \lambda_k = 1 ~~(0 \le \lambda_k \le 1),
\end{equation}
where $n_k$ is the density corresponding to degenerate ground state $\Psi_k$, that are not in $\bf{D}$~\cite{L83,L82}.

A convex set can be obtained by looking at ensembles. The density of an ensemble can be defined through the (statistical, or von Neuman) density operator 
\begin{equation}
\label{DD}
\hat{D} = 
\sum_{k=1}^{M} \lambda_k \ket{\Psi_k}\bra{\Psi_k},
~~\mbox{with} ~\sum_{k=1}^{M} \lambda_k = 1 ~~(0 \le \lambda_k \le 1)\;.
\end{equation}
The expectation value of an operator $\hat{O}$ on an ensemble is defined as
\begin{equation}\label{DO1}
O := \mbox{Tr} \left\{ \hat{D} \hat{O} \right\},
\end{equation}
where the symbol ``Tr'' stands for the trace taken over an arbitrary, complete set of orthonormal $N$-particle states
\begin{equation}\label{DO2}
\mbox{Tr}\{\hat{D}\hat{O}\} := \sum_{k}  \bra{\Phi_k}(\hat{D}\hat{O})\ket{\Phi_k}.
\end{equation}
The trace is invariant under unitary transformations of the complete set for the ground-state manifold of the Hamiltonian $\hat{H}$ [see Eq.(\ref{DD})]. Since
\begin{equation}
\label{TRDO}
\mbox{Tr} \left\{ \hat{D}\hat{O} \right\}  = \sum_{k=1}^{M} \lambda_k \bra{\Psi_k} \hat{O} \ket{\Psi_k},
\end{equation}
the energy obtained from a density matrix of the form (\ref{DD})
is the total ground-state energy of the system.

Densities of the form (\ref{ED}) are called ensemble v-representable densities, or E-V-densities. We denote this set of densities as $\bf{D_{EV}}$.
Densities that can be obtained from a single wavefunction are said to be pure-state (PS) v-representable, or PS-V-densities. The functional $F_{\rm HK}$ can then be extended as~\cite{VALONE:80}
\begin{equation}
F_{\rm EHK}[n] :=  Tr\left\{ \hat{D} \left(  \hat{T} + \hat{V}_{ee} \right) \right\},~~ \mbox{with}~
n \in \bf{D_{EV}}
\end{equation}
where $\hat{D}$ has the form (\ref{DD}) and is any density matrix giving the density $n$. However, the set $\bf{D_{EV}}$, just like $\bf{D}$, is difficult to characterize. Moreover, as for $F_{\rm HK}$ and $F_{\rm LL}$, a proof of the differentiability of $F_{\rm EHK}$ (and for the non-interacting versions of the same functional) is not available.

In the Lieb formulation, however, differentiability can be 
addressed to some extent~\cite{L83,Englisch:84a,Englisch:84b} .
In the work of Lieb, $\bf{P}$ is restricted to $\bf{P}={\cal L}^{3/2}(\mathbb{R}^3)+{\cal L}^\infty(\mathbb{R}^3)$ and
wavefunctions are required to be in
\begin{equation}
{\bf W_N}=\left\{\Psi\mid ||\Psi||=1, T[\Psi]\leq\infty\right\}.
\end{equation}

The universal functional is defined as 
\begin{equation}
\label{FL}
F_{\rm L}[n] := \inf_{\hat{D} \rightarrow n \in \bf{D_N}} 
Tr\left\{ \hat{D} \left(  \hat{T} + \hat{V}_{ee} \right) \right\},
\end{equation}
and it can be shown that the infimum is a minimum~\cite{L83}. Note that in definition (\ref{FL}), $\hat{D}$ is a generic density matrix of the form
\begin{equation}
\hat{D} = 
\sum_{k} \lambda_k \ket{\Psi_k}\bra{\Psi_k},
~~\mbox{with} ~\sum_{k} \lambda_k = 1 ~~(0 \le \lambda_k \le 1)~,
\end{equation}
where $\Psi_k \in \bf{W_N}$. The sum is not restricted to a finite number of degenerate ground states as in Eq. (\ref{DD}). This minimization over a larger, less restricted set leads to the statements
\begin{equation}
F_{\rm L}[n] \le F_{\rm LL}[n],~~\mbox{for}~ n \in \bf{D_N},
\end{equation}
and
\begin{equation}
F_{\rm L}[n] = F_{\rm LL}[n] = F_{\rm HK}[n],~~\mbox{for}~ n \in \bf{D}\;.
\end{equation}
$F_{\rm L}[n]$ is defined on a convex set, and it is a convex functional. 
This implies that $F_{\rm L}[n]$ is differentiable at any ensemble v-representable densities and nowhere else~\cite{L83}.
Minimizing the functional 
\begin{equation}
E_{\rm L}[n] :=  F_{\rm L}[n] + \int {d^3 r}~n({\bf r})v({\bf r})\;
\end{equation}
with respect to the elements of $\bf{D_{EV}}$ by the Euler-Lagrange equation
\begin{equation}
\frac{\delta F_{\rm L}}{\delta n({\bf r})} + v({\bf r}) = 0
\end{equation}
is therefore well-defined on the set $\bf{D_{EV}}$ and generates a valid energy minimum.

We finally address, although only briefly, some important points about the Kohn-Sham scheme and its rigorous justification. 
The results for $F_{\rm L}$ carry over to $T_{\rm L}[n]$. That is, the functional
\begin{equation}
T_{\rm L}[n] = \inf_{\hat{D} \rightarrow n} Tr \left\{ \hat{D} \hat{T}\right\},~~ \mbox{with}~n \in \bf{D_N}
\end{equation}
is differentiable at any non-interacting ensemble v-representable densities and nowhere else. We can gather all these densities in the set $\bf{D^{s}_{EV}}$. Then, the Euler-Lagrange equation
\begin{equation}
\frac{\delta T_{\rm L}}{\delta n({\bf r})} + v_{s}({\bf r}) = 0
\end{equation}
is well defined on the set $\bf{D^{s}_{EV}}$ only.  One can then redefine
the exchange-correlation functional as
\begin{equation}
\label{DEX2}
E_{xc,\rm L}[n] = F_{\rm L}[n] - T_{\rm L}[n] - U[n],
\end{equation}
and observe that the differentiability of $F_{\rm L}[n]$ and $T_{\rm L}[n]$
implies the differentiability of $E_{xc}[n]$ only on $\bf{D_{EV}} \cap \bf{D^{s}_{EV}}$.
The question as to the size of the latter set remains. For densities defined on a discrete lattice (finite or infinite) it is known~\cite{Chayes:85} that $\bf{D_{EV}} = \bf{D^{s}_{EV}}$. Moreover, in the continuum limit, $\bf{D_E}$ and $\bf{D^{s}_E}$ can be shown to be dense with respect to one another 
~\cite{L83,Englisch:84a,Englisch:84b}. This implies that  
any element of $\bf{D_{EV}}$ can be approximated, with an arbitrary accuracy,
by an element of $\bf{D^{s}_{EV}}$. But, whether or not the two sets coincide remains an open question.

\sec{Functional Approximations}
\label{sec:3}
Numerous approximations to $E_{xc}$ exist, each with its own successes and failures~\cite{B12}.  The simplest is the local density approximation (LDA), which had early success with solids\cite{KS65}.  LDA assumes that the exchange-correlation  energy density can be approximated locally with that of the 
uniform gas. DFT's popularity in the chemistry community skyrocketed upon development of the generalized gradient approximation (GGA)~\cite{P86}. Inclusion of density gradient dependence generated sufficiently accurate results to be useful in many chemical  and materials applications. 

Today, many scientists use hybrid 
functionals, which substitute a fraction of single-determinant exchange for part of the GGA exchange~\cite{B88, B93, LYP88}.  More recent developments in functional approximations include meta-GGAs~\cite{PS01}, 
which include dependence on the kinetic energy density, and hyper-GGAs~\cite{PS01}, which include exact exchange as input to the functional.  Inclusion of occupied and then unoccupied orbitals as inputs to functionals increases their complexity and computational cost; the idea that this increase is coupled with an increase in accuracy was compared to Jacob's Ladder~\cite{PS01}. The best approximations are based on the exchange-correlation hole, such as the real space cutoff of the LDA hole that ultimately led to the GGA called PBE~\cite{PBE96, PBE98}.  An introduction to this and
some other exact properties of the functionals  follows in the remainder of this section.

Another area of functional development of particular importance to the warm dense matter community is focused on orbital-free functionals~\cite{DG90,KJTH09,KJTH09b,KT12,WC00}.  These approximations bypass solution of the Kohn-Sham equations by directly approximating the non-interacting kinetic energy.  In this way, they recall the original, pure DFT of Thomas-Fermi theory~\cite{T27,F27,F28}. While many approaches have been tried over the decades, including fitting techniques from computer science~\cite{SRHM12}, no general-purpose solution of sufficient accuracy has been found yet.

\ssec{Exact Conditions}
Though we do not know the exact functional form for the universal functional, we do know some facts about its behavior and the relationships between its components.  
Collections of these facts are called exact conditions.  Some can be found by inspection of the formal definitions of the functionals and their variational properties.  
The correlation energy and its constituents are differences between functionals evaluated on the true and Kohn-Sham systems. As an example, consider the kinetic correlation:

\begin{equation}
T_c[n]=T[n]-T_s[n].
\end{equation}
Since the Kohn-Sham kinetic energy is the lowest kinetic energy of any wavefunction with density $n({\bf r})$, we know $T_c$ must be non-negative.  
Other inequalities follow similarly, as well as one from noting that the exchange functional is (by construction) never positive~\cite{B07}:

\begin{equation}
E_x\leq0,~E_c\leq0,~U_c\leq0,~T_c\geq0.
\end{equation}

Some further useful exact conditions are found by uniform coordinate scaling~\cite{LP85}.  
In the ground state, this procedure requires scaling all the coordinates of the wavefunction\footnote{Here and in the remainder of the chapter, we restrict ourselves to square-integrable wavefunctions over the domain $\mathbb{R}^{3N}$.} by a positive constant $\gamma$, while preserving normalization to $N$ particles:

\begin{equation}
\label{wfnscale}
\Psi_\gamma({\bf r_1},{\bf r_2},...,{\bf r_N})=\gamma^{3N/2}\Psi(\gamma{\bf r_1},\gamma{\bf r_2},...,\gamma{\bf r_N}),
\end{equation}
which has a scaled density defined as
\begin{equation}
\label{denscale}
n_\gamma({\bf r})=\gamma^3 n(\gamma {\bf r}).
\end{equation}
Scaling by a factor larger than one can be thought of as squeezing the density, while scaling by $\gamma<1$ spreads the density out. 
For more details on the many conditions that can be extracted using this technique and how they can be used in functional approximations, see Ref.~\cite{B07}.

Of greatest interest in our context are conditions involving exchange-correlation and other components of the universal functional. 
Through application of the foregoing definition of uniform scaling, we can write down some simple uniform scaling equalities.  Scaling the density yields
\begin{equation}
T_s[n_\gamma]=\gamma^2 T_s[n]
\end{equation}
for the non-interacting kinetic energy and
\begin{equation}
E_x[n_\gamma]=\gamma~E_x[n]
\end{equation}
for the exchange energy.  Such simple conditions arise because these functionals are defined on the non-interacting Kohn-Sham Slater determinant.  
On the other hand, although the density from a scaled interacting wavefunction is the scaled density, the scaled wavefunction is not the ground-state wavefunction of the scaled density. This means correlation scales less simply and only inequalities can be derived for it.

Another type of scaling that is simply related to coordinate scaling is interaction scaling, the adiabatic change of the interaction strength~\cite{PK03}.  In the latter, the electron-electron interaction in the Hamiltonian, $V_{ee}$, is multiplied by a factor, $\lambda$ between 0 and 1, while holding $n$ fixed.  When $\lambda=0$, interaction vanishes.  At $\lambda=1$, we return to the Hamiltonian for the fully interacting system.  Due to the simple, linear scaling of $V_{ee}$ with coordinate scaling, we can relate it to scaling of interaction strength. Combining this idea with some of the simple equalities above leads to one of the most powerful relations in ground-state functional development, the adiabatic connection formula~\cite{GL76,LP75}:
\begin{equation}
E_{xc}[n]=\int_0^1 d\lambda U_{xc}[n](\lambda),
\label{acform}
\end{equation}
where
\begin{equation}
U_{xc}[n](\lambda)=V_{ee}[\Psi^{\lambda}[n]]-U[n]
\end{equation}
and $\Psi^{\lambda}[n]$ is the ground-state wavefunction of density $n$ for a given $\lambda$ and
\begin{equation}
\Psi^{\lambda}[n]({\bf r_1},{\bf r_2},...,{\bf r_N})=\lambda^{3N/2}\Psi[n_{1/\lambda}](\lambda{\bf r_1},\lambda{\bf r_2},...,\lambda{\bf r_N}).
\end{equation}

Interaction scaling also leads to some of the most important exact conditions for construction of functional approximations, the best of which are based on the exchange-correlation hole.  The exchange-correlation hole represents an important effect of an electron sitting at a given position. All other electrons will be kept away from this position by exchange and correlation effects, due to the antisymmetry requirement and the Coulomb repulsion, respectively. This representation allows us to calculate $V_{ee}$, the electron-electron repulsion, in terms of an electron distribution function.\footnote{For a more extended discussion of these topics, see Ref.~\cite{PK03}.} 

\def\br{{\bf r}}
To define the hole distribution function, we need first to introduce the
pair density function. The pair density, $P({\bf r},{\bf r'})$ describes the distribution of the electron pairs. This is proportional to the the probability of
finding an electron in a volume $d^3r$ around position ${\bf r}$ {\it and}
a second electron in the volume $d^3r'$ around ${\bf r'}$.  
In terms of the electronic wavefunction, it is written as follows
\begin{equation}
\label{pairden}
P({\bf r},{\bf r'})=N(N-1)\int d^3 r_3 \ldots \int d^3 r_N~| \Psi({\bf r},{\bf r'},\ldots ,{\bf r}_N)|^{2}.
\end{equation}
We then can define the conditional probability density of finding 
an electron in $d^3r'$ after having already found one at ${\bf r}$, which 
we will denote $n_2({\bf r},{\bf r'})$. Thus
\begin{equation}
n_2({\bf r},{\bf r'})=
P({\bf r},{\bf r'})/n({\bf r}).
\end{equation}
If the positions of the electrons were truly independent of one another (no electron-electron interaction and no antisymmetry requirement
for the wavefunction) this would be just $\n(\br')$, independent of ${\bf r}$.  
But this cannot be, as 
\begin{equation}
\int d^3r'~n_2({\bf r},{\bf r'})=N-1.
\end{equation}
The conditional density integrates to one fewer electron, since one electron is at the reference point.  
We therefore define a ``hole" density:
\begin{equation}
n_2({\bf r},{\bf r'})=n({\bf r'})+n_{\rm hole}({\bf r},{\bf r'}).
\end{equation}
which is typically negative and integrates to -1
~\cite{PK03}:
\begin{equation}
\int d^3 r'~n_{\rm hole}({\bf r},{\bf r'})=-1.
\end{equation}
The exchange-correlation hole in DFT is given by the coupling-constant average:
\begin{equation}
n_{xc}({\bf r},{\bf r'})=\int_0^1 d\lambda~n^\lambda_{\rm hole}({\bf r},{\bf r'}),
\end{equation}
where $n^\lambda_{\rm hole}$ is the hole in $\Psi^{\lambda}$. 
So, via the adiabatic connection formula (Eq.~\ref{acform}), the exchange-correlation energy can be written 
as a double integral over the exchange-correlation hole:
\begin{equation}
E_{xc}=\frac{1}{2}\int d^3r~ n({\bf r})\int d^3 r'~\frac{n_{xc}({\bf r},{\bf r'})}{\left|{\bf r}-{\bf r'}\right|}.
\end{equation}

By definition, the exchange hole is given by $n_{x} = n^{\lambda = 0}_{\rm hole}$ 
and the correlation hole, $n_{c}$, is everything {\em not} in $n_{x}$.
The exchange hole may be readily obtained from the (ground-state) pair-correlation function of the Kohn-Sham system.
Moreover $n_{x}({\bf r},{\bf r}) = 0$, $n_{x}({\bf r},{\bf r'}) \le 0$, and for one particle systems $n_{x}({\bf r},{\bf r'}) = - n({\bf r'})$.
If the Kohn-Sham state is a single Slater determinant, then the exchange energy assumes the form of the
Fock integral evaluated with occupied Kohn-Sham orbitals.
It is straightforward to verify that the exchange-hole satisfies the sum rule
\begin{equation}
\int d^3 r'~n_{x}({\bf r},{\bf r'})=-1\;;
\end{equation}
and thus
\begin{align}
\int d^3 r'~n_{c}({\bf r},{\bf r'}) =0\;.
\end{align}
The correlation hole is a more complicated quantity, and its contributions oscillate from negative to positive in sign. 
Both the exchange and the correlation hole decay to zero at large distances from the reference position ${\bf r}$. 

These and other conditions on the exact hole are used to constrain exchange-correlation functional approximations.  The seemingly unreasonable
reliability of the simple LDA has been explained as the result of the ``correctness" of the LDA exchange-correlation hole~\cite{EBPb96,JG89}.  Since the LDA is constructed from the uniform gas, which has many realistic properties, its hole satisfies many mathematical conditions on this quantity~\cite{BPL94}. Many of the most popular improvements on LDA, including the PBE generalized gradient approximation, are based on models of the exchange-correlation hole, not just fits of exact conditions or empirical data~\cite{PBE96}. In fact, the most successful approximations usually are based on models for the exchange-correlation hole, which can be explicitly tested~\cite{CFN00}.
Unfortunately, insights about the ground-state exchange-correlation hole do not simply generalize as temperatures increase, as will be discussed later.

\sec{Thermal DFT}
\label{sec:4}

Thermal DFT deals with statistical ensembles of quantum states describing the thermodynamical equilibrium of many-electron systems. 
The grand canonical ensemble is particularly convenient to deal with the symmetry of identical particles.
In the limit of vanishing temperature, thermal DFT reduces to an equiensemble  ground state DFT description~\cite{E10},
which, in turn, reduces to the standard pure-state approach for non-degenerate cases.

While in the ground-state problem the focus is on the ground state energy, in the statistical mechanical framework the focus is on the grand canonical potential.
Here, the grand canonical Hamiltonian plays an analogous role as the one played by the Hamiltonian for the ground-state problem. The former 
is written
\begin{equation}\label{gcop}
\hat{\Omega} = \hat{H}  - \tau \hat{S} - \mu \hat{N},
\end{equation}
where $\hat{H}$, $\hat{S}$, and $\hat{N}$ are the Hamiltonian, entropy, and particle-number operators. 
The crucial quantity by which the Hamiltonian differs from its grand-canonical version is the entropy operator:\footnote{Note that, we eventually 
choose to work in a system of units such that the Boltzmann
constant is $k_B = 1$, that is, temperature is measured in energy units.}
\begin{equation}\label{entropyop}
\hat{S} = - ~ k_B \ln \hat{\Gamma} \;,
\end{equation}
where 
\begin{equation}\label{statop}
\hat{\Gamma}= \sum_{N,i} {w_{N,i}} \ket{\Psi_{N,i}} \bra{\Psi_{N,i}}\;.
\end{equation}
$\ket{\Psi_{N,i}}$ are orthonormal 
$N$-particle states (that are not necessarily eigenstates in general)
and $w_{N,i}$ are normalized statistical weights satisfying $\sum_{N,i} w_{N,i} = 1$.  
$\hat{\Gamma}$ allows us to describe the thermal ensembles of interest.

Observables are obtained from the statistical average of Hermitian operators
\begin{equation}
O[\hat{\Gamma}] =  {\mbox{Tr}} ~ \{\hat{\Gamma}\hat{O}\} = \sum_N \sum_i w_{N,i} \bra{\Psi_{N,i}} \hat{O} \ket{\Psi_{N,i}} \; . 
\end{equation}
These expressions are similar to Eq.~(\ref{TRDO}), but here the trace is not restricted to the ground-state manifold.  

In particular, consider the average of the $\hat{\Omega}$, $\Omega[\hat{\Gamma}]$, and search for its minimum at a given temperature, $\tau$, 
and chemical potential, $\mu$.
The quantum version of the Gibbs Principle ensures that the minimum exists and is unique 
(we shall not discuss the possible complications introduced by  the occurrence of phase transitions).  
The minimizing statistical operator is the grand-canonical statistical operator, with statistical weights given by
\begin{equation}
w_{N,i}^0 = \frac{ \exp[-\beta({E}_{N,i}^0-\mu N)] }{\sum_{N,i}  \exp[-\beta({E}_{N,i}^0-\mu N)] }.
\end{equation} 
${E}_{N,i}^0$ are the eigenvalues of $N$-particle eigenstates.
It can be verified that  $\Omega[\hat{\Gamma}]$ may be written in the usual form
\begin{equation}
\Omega=E-\tau S-\mu N=-k_B\tau\ln{Z_G},
\end{equation}
where $Z_G$ is the grand canonical partition function; which is defined by
\begin{align}
Z_G&=\sum_N \sum_j e^{-\beta({E}_{N,i}^0-\mu N)}\;.
\end{align}

The statistical description we have outlined so far is the standard one. Now, we wish to switch to a density-based description and thereby enjoy the same benefits as in the ground-state problem.
To this end, the minimization of $\Omega$ can be written as follows:
\begin{equation}
\label{Mermin}
\Omega^\tau_{v-\text{\scriptsize{$\mu$}}} = \min_\n \left\{F^{\tau}[n] + \int d^3r~n({\bf r}) (v({\bf r})-\mu)
\right\}
\end{equation}
with $n({\bf r})$ an ensemble $N$-representable density and 
\begin{equation}
\F^{\tau}[n] :=  \min_{\hat{\Gamma}\to n}  F^{\tau} [\hat{\Gamma}]  = \min_{\hat{\Gamma}\to n} \left\{ T[\hat{\Gamma}] + V_{ee}[\hat{\Gamma}] - \tau S[\hat{\Gamma}]\right\}.
\label{Ft}
\end{equation}
This is the constrained-search analog of the Levy functional~\cite{L79,PY89}, Eq.~(\ref{LL}). It replaces the functional originally defined by Mermin~\cite{M65} in the same way that Eq.~(\ref{LL}) replaces Eq.~(\ref{FHK}) in the ground-state theory. 
\footnote{The interested reader may find the extension of the Hohenberg-Kohn theorem to the thermal framework in Mermin's paper.}

Eq.~(\ref{Ft}) defines the thermal universal functional. Universality of this quantity means that
{\em it does not depend explicitly on the external potential nor on $\mu$}.
This is very appealing, as it hints at the possibility of widely applicable approximations.

We identify ${\Gamma}^{\tau}[n]$ as the minimizing statistical operator in Eq.~(\ref{Ft}).
We can then define other interacting density functionals at a given temperature by taking the trace over the given minimizing statistical operator.
For example, we have:
\begin{align}
T^{\tau}[n]&:=T[\hat{\Gamma}^{\tau}[n]]\\
V_{ee}^{\tau}[n]&:=V_{ee}[\hat{\Gamma}^{\tau}[n]]\\
S^{\tau}[n]&:=S[\hat{\Gamma}^{\tau}[n]].
\end{align}

In order to introduce the thermal Kohn-Sham system, we proceed analogously as in the zero-temperature case.
We assume that there exists an ensemble of non-interacting systems with same average particle density {\em and} temperature of the interacting ensemble.
Ultimately, this determines the one-body Kohn-Sham potential, which includes the corresponding chemical potential.  
Thus, the noninteracting (or Kohn-Sham) universal functional is defined as 
\begin{equation}
F_{s}^{\tau}[n] :=  \min_{\hat{\Gamma}\to n} K^{\tau}[\hat{\Gamma}] = K^{\tau}[\hat{\Gamma}_s^{\tau}[n]]=K_s^\tau[n]
\label{Fst},
\end{equation}
where  $\hat{\Gamma}_s^{\tau}[n]$ is a statistical operator that describes the Kohn-Sham ensemble and
$K^{\tau}[\hat{\Gamma}] := T[\hat{\Gamma}] - \tau S[\hat{\Gamma}]$ is a combination we have chosen to call the kentropy.

We can also write the corresponding Kohn-Sham equations at non-zero temperature, which are analogous to Eqs.~(\ref{KS1}) and~(\ref{VS})~\cite{KS65}:
\begin{equation}
\label{FTKS1}
\left[-\frac{1}{2}
\nabla^2+v_s({\bf r})\right] \varphi_{i}({\bf r})= \epsilon^{\tau}_{i} \varphi_{i}({\bf r})
\end{equation}
\begin{equation}
\label{FTKS2}
v_s({\bf r}) = v_{H}[n]({\bf r}) + v_{xc}[n]({\bf r}) + v({\bf r}).
\end{equation}
The accompanying density formula is
\begin{equation}
n({\bf r})=\sum_i f_i |\varphi_{i}({\bf r})|^2,
\end{equation}
where 
\begin{equation}
f_i=\left(1+e^{\left(\epsilon^{\tau}_i-\mu\right)/\tau}\right)^{-1}.
\end{equation} 
Eqs.~(\ref{FTKS1}) and~(\ref{FTKS2}) look strikingly similar to the case of non-interacting
Fermions. However, the Kohn-Sham weights, $f_i$, are not simply the familiar Fermi functions, due to the temperature dependence of the Kohn-Sham eigenvalues.

Through the series of equalities in Eq.~(\ref{Fst}), we see that the non-interacting universal density functional
is obtained by evaluating the kentropy on a non-interacting, minimizing statistical operator which, at temperature $\tau$, yields the
average particle density $n$. The seemingly simple notation of Eq.~(\ref{Fst}) reduces the kentropy first introduced  as a functional of the statistical operator
to a finite-temperature functional of the density. 
From the same expression, we see that the kentropy plays a role in this framework analogous to that of the kinetic 
energy within ground-state DFT. Finally, we spell-out the components of $F_s^\tau[n]$:
\begin{equation}
F_s^\tau[n]=T_s^\tau[n]-\tau S_s^\tau[n]\;,
\end{equation}
where $T_s^\tau[n] := T[\hat{\Gamma}_s^{\tau}[n]]$ and $S_s^\tau[n] := S[\hat{\Gamma}_s^{\tau}[n]]$.

Now we identify other fundamental thermal DFT quantities. 
First, consider the decomposition of the interacting grand-canonical potential as a functional of the density given by
\begin{equation}
\Omega^\tau_{v-\text{\scriptsize{$\mu$}}}[n] = F_s^\tau[n] + U[n] + {\cal F}^\tau_{xc}[n] + \int d^3r~n({\bf r}) \left( v({\bf r})-\mu \right)\;.
\end{equation}
Here, $U[n]$ is the Hartree energy having the form in Eq.~(\ref{UH}). The adopted notation stresses that temperature dependence of $U[n]$ enters only through the input equilibrium density.  The exchange-correlation free-energy density functional
is given by
\begin{equation}
 {\cal F}^\tau_{xc}[n] =  F^\tau[n] - F_s^\tau[n] - U[n]\;.
\end{equation}
It is also useful to introduce a further decomposition:
\begin{equation}
{\cal F}_{xc}^\tau[n] := {\cal F}_x^\tau[n] + {\cal F}_c^\tau[n]\;.
\end{equation}
This lets us analyze the two terms on the right hand side along with their components. 

The exchange contribution is 
\begin{equation}\label{Ftx}
 {\cal F}^\tau_{x}[n] = V_{\rm ee}[\Gamma^\tau_s[n]] -  U[n]\;.
\end{equation}
Note that the average on the right hand side is taken with respect to
the Kohn-Sham ensemble and that kinetic and entropic contributions do not contribute to exchange effects explicitly.
Interaction enters in Eq.~(\ref{Ftx}) in a fashion that is reminiscent of (but not the same as) finite-temperature Hartree-Fock  theory. 
In fact, ${\cal F}^\tau_{x}[n]$ may be expressed in terms of the square modulus of the finite-temperature Kohn-Sham one-body density matrix.
%~\cite{ED11}.
Thus ${\cal F}^\tau_{x}[n]$ has an explicit, known expression, just as does ${\cal F}^\tau_{s}[n]$. 
For the sake of practical calculations, however, approximations are still needed. 

The fundamental theorems of density functional theory were
proven for any ensemble with monotonically decreasing
weights~\cite{T79} and were applied to extract excitations~\cite{GOK88, OGKb88, N98}. But simple approximations to the exchange for such ensembles are
corrupted by ghost interactions~\cite{GPG02}
contained in the ensemble Hartree term.  
The Hartree energy defined in Eq.~(\ref{UH}) is defined as the electrostatic self-energy of the density, both
for ground-state DFT and at non-zero temperatures.  But the physical ensemble of Hartree energies
is in fact the Hartree energy of each ensemble member's density, added together with the weights
of the ensembles.  Because the Hartree energy is quadratic in the density, it therefore
contains ghost interactions~\cite{GPG02}, i.e., cross terms, that are unphysical.  These must
be canceled by 
the exchange energy, which must therefore contain a contribution:
\begin{equation}
\Delta E^{GI}_X= \sum_i w_i U[n_i] - U\left[\sum_i w_i n_i\right].
\end{equation}
Such terms appear only when the temperature is non-zero and so are missed by 
any ground-state approximation to $E_x$.

Consider, now, thermal DFT correlations. 
We may expect correctly that these will be obtained as differences between interacting averages and the noninteracting ones.
The kinetic correlation energy density functional is
\begin{equation}
T_c^\tau[n] :=  T^{\tau}[n]-T_s^\tau[n],
\end{equation}
and similar forms apply to $S_c^\tau[n]$ and $K_c^\tau[n]$. 
Another important quantity is the correlation potential density functional. At finite-temperature, this
is defined by
\begin{equation}
U^{\tau}_c[n] :=  V_{\rm ee}[\Gamma^{\tau}[n]] - V_{\rm ee}[\Gamma^{\tau}_s[n]]\;.
\end{equation}

Finally, we can write the correlation free energy as follows
\begin{equation}\label{Ftc}
{\cal F}^\tau_{c}[n]=K_c^\tau[n]+U^\tau_c[n] =  E_c^\tau[n] - \tau S_c^\tau[n]
\end{equation}
where
\begin{equation}
K_c^\tau[n] = T_c^\tau[n] - \tau S^\tau_c[n]\;
\end{equation}
is the correlation kentropy density functional and
\begin{equation}
E_c^\tau[n] := T_c^\tau[n] + U_c^\tau[n]\;
\end{equation}
generalizes the expression of the correlation energy to finite temperature.
Above, we have noticed that entropic contributions do not enter explicitly in the definition of ${\cal F}^\tau_x[n]$.
From Eq.~(\ref{Ftc}), on the other hand, we see that the correlation entropy is essential for determining  ${\cal F}^\tau_c[n]$. Further, it may
be grouped together the kinetic contributions (as in the first identity) or separately (as in the second identity), depending on the context of the current analysis.

In the next section, we consider finite-temperature analogs of the exact conditions described earlier for the ground state functionals. 
This allow us to gain additional insights about the quantities identified so far.

\sec{Exact Conditions at Non-Zero Temperature}
\label{sec:5}

In the following, we review several properties of the basic energy components of thermal Kohn-Sham DFT~\cite{PPFS11,DT11}.

We start with some of the most elementary properties, their signs~\cite{PPFS11}:
\begin{equation}
{\cal F}_x^\tau[n] \leq 0,~ {\cal F}_c^\tau[n] \leq 0,~ U_c^\tau[n] \leq 0,~ K_c^\tau[n] \geq 0.
\end{equation}
The sign of  ${\cal F}_x^\tau[n]$ is evident from the definition given in terms of the Kohn-Sham one-body reduced density matrix~\cite{ED11}.  
The others may be understood in terms of their variational properties.
For example, let us consider the case for $K_c^\tau[n]$. We know that the Kohn-Sham statistical operator minimizes the kentropy
\begin{equation}
K_s^\tau[n]=K^\tau[\hat{\Gamma}_s^\tau[n]]\;.
\end{equation}
Thus, we also know that $K_s^\tau[n]$ must be less than $K^\tau[n] = K^\tau[\Gamma^\tau[n]]$, where $\Gamma^\tau[n]$ is the
equilibrium statistical operator. This readily implies that
\begin{equation}
K_c^\tau[n] = K^\tau[\hat{\Gamma}^\tau[n]] - K^\tau[\hat{\Gamma}_s^\tau[n]] \geq 0.
\end{equation}
An approximation for $K_c^\tau[n]$ that does not respect this inequality will
not simply have the ``wrong" sign. Much worse is that results from such an approximation will suffer from improper variational character.

A set of remarkable and useful properties are the scaling relationships. 
What follows mirrors the zero-temperature case, but an important and intriguing difference is  
the relationship between coordinate and temperature scaling. 

We first introduce the concept of uniform scaling of statistical ensembles in terms of a particular scaling of the corresponding statistical operators.
 \footnote{Uniform coordinate scaling may be considered as (very) careful dimensional analysis applied to density functionals. Dufty and Trickey analyze
non-interacting functionals in this way in Ref.~\cite{DT11}.}
Wavefunctions of each state in the ensemble can be scaled as in Eq. (\ref{wfnscale}). At the same time, we require that the statistical mixing is not affected, so the statistical weights are held fixed under scaling (we shall return to this point in Section \ref{ssec:4.1}).
In summary, the scaled statistical operator is
\begin{equation}
\label{gamscale}
\hat{\Gamma}_\gamma := \sum_N\sum_i w_{N,i}\ket{\Psi_{\gamma,N,i}}\bra{\Psi_{\gamma,N,i}},
\end{equation}
where (the representation free) Hilbert space element $\ket{\Psi_\gamma}$ is such that
$\Psi_{\gamma}({\bf r}_1,...,{\bf r}_N)=\bra{{\bf r}_1,...,{\bf r}_N}\Psi_\gamma\rangle$.
For sake of simplicity, we restrict ourselves to states of the type typically considered in the
ground-state formalism.  

Eq.~(\ref{gamscale}) leads directly to scaling relationships for any observable.  
For instance, we find
\begin{align}
N[\hat{\Gamma}_\gamma]&=N[\hat{\Gamma}],\\
T[\hat{\Gamma}_\gamma]&=\gamma^2 T[\hat{\Gamma}], \text{ and}\\
S[\hat{\Gamma}_\gamma]&=S[\hat{\Gamma}]\;.
\end{align}
Combining these, we find
\begin{equation}\label{Fsscaling}
\hat{\Gamma}_s^\tau[n_\gamma]=\hat{\Gamma}_{\gamma,s}^{\tau/\gamma^2}[n] \text{ and }
F_s^{\tau}[n_\gamma]= \gamma^2 F_s^{\tau/\gamma^2}[n].
\end{equation}
Eq.~(\ref{Fsscaling}) states that the value of the non-interacting 
universal functional evaluated at a scaled density is related to the
value of the same functional evaluated on the unscaled density at a scaled temperature.
Eq.~(\ref{Fsscaling}) constitutes a powerful statement, which becomes more apparent by rewriting it as follows~\cite{PPFS11}:
\begin{equation}
F_s^{\tau'}[n]= \frac{\tau'}{\tau} F_s^\tau[\n_{{\sqrt{\tau/\tau'}}}].
\end{equation}  
This means that,  if we know $F_s^{\tau}[n]$ at some non-zero temperature $\tau$, we can find its value at any other temperature by scaling its argument.

Scaling arguments allow us to extract other properties of the functionals, such as some of their limiting behaviors.  
For instance, we can show that in the ``high-density" limit, the kinetic term dominates~\cite{PPFS11}:
\begin{equation}
T_s^{\infty}[n]=\lim_{\gamma \rightarrow +\infty} F_s[n_\gamma]/\tau^2\;
\end{equation}
while in the ``low-density" limit, the entropic term dominates:
\begin{equation}
S_s^{\infty}[n]=\lim_{\gamma \rightarrow 0}F_s[n_\gamma]\tau.
\end{equation}

Also, we may consider the interacting universal functional for a system with coupling strength equal to  $\lambda$
\begin{equation}
F^{\tau,\lambda}[n]=\min_{\hat{\Gamma}\rightarrow n}\left\{T[\hat{\Gamma}]+\lambda V_{ee}[\hat{\Gamma}]-\tau S[\hat{\Gamma}]\right\}, 
\end{equation}
and note that in general,
\begin{equation}
\hat{\Gamma}^{\tau,\lambda}[n]\neq\hat{\Gamma}^{\tau}[n].
\end{equation}
We can relate these two statistical operators~\cite{PPFS11}. In fact, one can prove
\begin{equation}
\label{funcscale}
\hat{\Gamma}^{\tau,\lambda}[n]=\hat{\Gamma}_\lambda^{\tau/\lambda^2}[n_{1/\lambda}] \text{ and }
F^{\tau,\lambda}[n]=\lambda^2 F^{\tau/\lambda^2}[n_{1/\lambda}].
\end{equation}
In the expressions above, a single superscript implies full interaction~\cite{PPFS11}. 
Eq.~(\ref{funcscale}) demands scaling of the coordinates, the temperature, and the
strength of the interaction at once. This procedure connects one equilibrium state to another equilibrium state, that of a ``scaled" system. 
Eq.~(\ref{funcscale}) may be used to state other relations similar to those discussed above for the
non-interacting case.

Scaling relations combined with the Hellmann-Feynman theorem allow us to generate the thermal analog of one of the most important
 statements of ground-state DFT, the adiabatic connection formula~\cite{PPFS11}:
\begin{equation}
{\cal F}_{xc}^{\tau}[n] = \int_0^1 d
\lambda ~ U_{xc}^\tau[n](\lambda),
\label{ACF}
\end{equation}
where
\begin{equation}
U_{xc}^\tau[n](\lambda)=V_{ee}[\hat{\Gamma}^{\tau,\lambda}[n]]-U[n]
\end{equation}
and a superscript $\lambda$ implies an electron-electron interaction strength equal to $\lambda$.
The interaction strength runs between zero, corresponding to the noninteracting Kohn-Sham system, and one, which 
gives the fully interacting system. All this must be done while keeping the density constant.
In thermal DFT, an expression like Eq.~(\ref{ACF}) offers the appealing possibility of defining an approximation for ${\cal F}_{xc}^{\tau}[n]$ {\em without} having to deal with kentropic contributions explicitly.

Another interesting relation generated by scaling connects the exchange-correlation to the exchange-only free energy~\cite{PPFS11}:
\begin{equation}
{\cal F}_x^\tau[n] =\lim_{\gamma\to\infty} 
{\cal F}_{xc}^{\gamma^2\tau}[n_\gamma]/\gamma.
\label{Exlimdef}
\end{equation}
This may be considered the definition of the exchange contribution in an xc functional, and so Eq.~(\ref{Exlimdef}) may also be used to extract an approximation for the exchange free energy, if  an approximation for the exchange-correlation 
free energy as a whole is given (for example, if obtained from Eq.~(\ref{ACF})).

Despite decades of research~\cite{P79b,PD84,PD00,DAC86}, thermal exchange-correlation GGAs have not been fully developed.
The majority of the applications in the literatures have adopted two {\em practical} methods: 
one uses plain finite-temperature LDA, the other uses ground-state GGAs within the thermal Kohn-Sham scheme.
This latter method ignores any modification to the exchange-correlation free energy functional due to its non-trivial temperature dependence.
As new approximations are developed, exact conditions such as those above are needed to define consistent 
and reliable thermal approximations.

\sec{Discussion}
\label{sec:6}

In this section, we discuss  several aspects that may not have been fully clarified by the previous, relatively abstract sections.
First,  by making use of a simple example, we will illustrate in more detail the tie between temperature and coordinate scaling.
Then, with the help of another example, we will show how scaling and other exact properties of the functionals can guide development and understanding of approximations.
The last subsection notes some complications in importing tools directly from ground-state methods to thermal DFT.  

\ssec{Temperature and Coordinate Scaling}
\label{ssec:4.1}

Here we give an illustration of how the scaling of the statistical operators introduced in the previous section is applicable to
thermal ensembles.   Our argument applies  -- with proper modifications and additions, such as the scaling of the
interaction strength -- to all Coulomb-interacting systems with all one-body external potentials. For sake of simplicity,
we shall restrict ourselves to non-interacting fermions in a one-dimensional harmonic oscillator at thermodynamic equilibrium. 

Let us start from the general expression of the Fermi occupation numbers 
\begin{equation}
n_i(\tau,\mu\epsilon_i)=\left(1+e^{\beta(\epsilon_i-\mu)}\right)^{-1},
\end{equation}
where $\epsilon_i$ is the $i^{\rm th}$ eigenvalue of the harmonic oscillator, $\epsilon_i=\omega(i+1/2)$.
For our system, the (time-independent) Schr\"{o}dinger equation is:
\begin{equation}
\left\{-\frac{1}{2}\frac{d^2}{dx^2}+v(x)\right\}\phi_i(x)=\epsilon_i \phi_i(x)\;.
\end{equation}
Now, we multiply the $x$-coordinates by $\gamma$
\begin{equation}
\left\{-\frac{1}{2\gamma^2}\frac{d^2}{dx^2}+v(\gamma x)\right\}\phi_i(\gamma x)=\epsilon_i \phi_i(\gamma x).
\end{equation}
We then multiply both sides by $\gamma^2$:
\begin{equation}
\left\{-\frac{1}{2}\frac{d^2}{dx^2}+\gamma^2v(\gamma x)\right\}\phi_i(\gamma x)=\gamma^2 \epsilon_i \phi_i(\gamma x).
\end{equation}
Substituting $\tilde{v}(x) = \gamma^2 v(\gamma x)$,  $\tilde{\phi}_i(x)= \sqrt{\gamma} \phi_i(\gamma x)$ (to maintain normalization), 
and $\tilde{\epsilon}_i=\gamma^2 \epsilon_i$ yields
\begin{equation}\label{Scaled}
\left\{-\frac{1}{2}\frac{d^2}{dx^2}+ \tilde{v}(x)\right\}\tilde{\phi}_i(x)=\tilde{\epsilon}_i \tilde{\phi}_i(x).
\end{equation}
The latter may be interpreted as the Schr\"{o}dinger equation for a ``scaled" system.
In the special case of the harmonic oscillator,
\begin{equation}
\gamma^2 v( \gamma x)=\gamma^4 v(x),
\end{equation}
the frequency scales quadratically, consistent with the scaling of the energies described just above.
Now, let us look at the occupation numbers for the ``scaled" system
\begin{equation}
n_i(\tau,\tilde{\mu},\tilde{\epsilon}_i)=\left(1+e^{\beta(\tilde{\epsilon}_i-\tilde{\mu})}\right)^{-1},
\end{equation}
where $\tilde{\mu}=\gamma^2\mu$ (in this way, the average number of particle is kept fixed too).  
These occupation numbers are equal to those of the original system at a temperature $\tau/\gamma^2$,
\begin{equation}
n_i(\tau,\tilde{\mu},\tilde{\epsilon}_i)=n_i(\tau/\gamma^2,\mu,\epsilon_i).
\end{equation}
Thus the statistical weights of the scaled system are precisely those of the original system, at a suitably scaled temperature.

\ssec{Thermal-LDA for Exchange Energies}
In ground-state DFT, uniform coordinate scaling of the exchange has been used to constrain the form of the exchange-enhancement factor in GGAs.
In thermal DFT, a ``reduction" factor, $R_x$, enters already in the expression of a LDA for the exchange energies. 
This lets us capture the reduction in exchange with increasing temperature, while 
keeping the zero-temperature contribution well-separated from the modification entirely due to non-vanishing temperatures.

The behavior of $R_x$ can be understood using the basic scaling relation for the exchange free energy.
Observe that, from the scaling of $\hat{\Gamma_s}^{\tau}$, $U$, and $V_{ee}[\Gamma^{\tau}_s[n]]$, one readily arrives at
\begin{equation}
\label{Fxng}
{\cal F}_x^\tau[n_\gamma]=\gamma {\cal F}_x^{\tau/\gamma^2}[n].
\end{equation}   
Since
\begin{equation}
{\cal F}_x^{{\rm LDA},\tau}[n]=\int~d^3 r~f_x^\tau(n(r)),
\end{equation}
Eq.~(\ref{Fxng}) implies that a thermal-LDA exchange free energy density must have the form~\cite{PPFS11}
\begin{equation}\label{fxlda}
f_x^{{\rm unif},\tau}(n)= e_x^{\rm unif}(n) R_x(\Theta),
\end{equation}
where $e_x^{\rm unif}(n) = - A_x n^{4/3}$, $A_x = (3/4 \pi)(3 \pi^2)^{1/3}$, and 
$R_x$ can only depend on $\tau$ and $n$ through the electron degeneracy   
$\Theta=2\tau/(3\pi^2 n({\bf r}))^{2/3}$.

The LDA is exact for the uniform electron gas and so automatically satisfies many conditions.
 As such, it also reduces to the ground-state LDA as temperature drops to zero:
\begin{equation}
R_x\rightarrow1~{\rm as}~\tau\rightarrow 0.
\end{equation}
Moreover, for fixed $n$, we expect 
\begin{equation}
{\cal F}^\tau_x/U\rightarrow0~{\rm as}~\tau\rightarrow \infty
\end{equation}
because the effect of the Pauli exclusion principle drops off as the behavior of the system becomes more classical.  
Moreover, since $U[n]$ does not depend explicitly on the temperature,  fixing $n$ also fixes $U$. 
We conclude that, the reduction factor must drop to zero:
\begin{equation}
R_x\rightarrow0~{\rm as}~\tau\rightarrow \infty.
\end{equation}

\begin{figure}[htbp]
\includegraphics[width=\columnwidth]{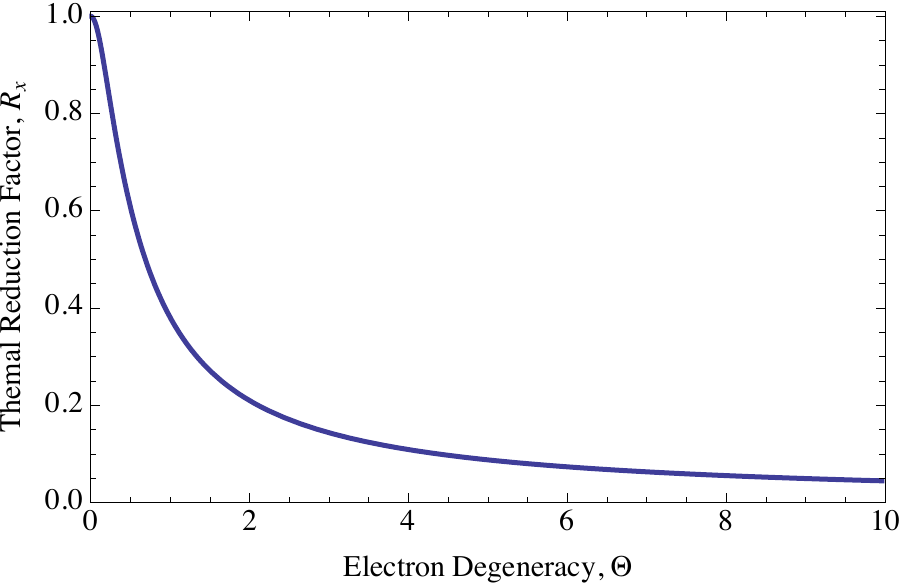}
\caption{Perrot and Dharma-Wardana's parameterization~\cite{PD84} of the thermal reduction factor for the exchange free energy for the uniform gas is plotted versus the electron degeneracy parameter \label{fig:tLDAx}
}\label{Rxfig}
\end{figure}

Now, let us consider the parameterization of $R_x$ for the uniform gas by Perrot and Dharma-Wardana~\cite{PD84}:
\begin{align}\label{Rxp}
R^{\rm unif}_x(\Theta)&\approx\notag\\
 \left(\frac{4}{3}\right)&\frac{0.75+3.04363\Theta^2-0.092270\Theta^3+1.70350\Theta^4}{1+8.31051\Theta^2+5.1105\Theta^4}\notag\\
&\times\tanh{ \Theta^{-1} }\;,
\end{align}
Here, $\Theta=\tau/\epsilon_F=2\tau/k_F^2$ and $k_F$ is the Fermi wavevector.  Note the factor of $4/3$ that is not present in their original paper, which arises because we include a factor of $3/4$ in $A_x$ they do not.  Fig. \ref{Rxfig} shows the plot of this reduction factor.
From both Fig. \ref{Rxfig} and Eq.~(\ref{Rxp}), it is apparent that the parametrization satisfies all the exact behaviors discussed just above.

\ssec{Exchange-Correlation Hole at Non-Zero Temperature}

Previously, we have emphasized that  in ground-state DFT, the exchange-correlation hole function was vital for constructing reliable  approximations.
Therefore, it is important to reconsider this quantity in the context of thermal DFT.  As we show below, this does not come without surprises.

In the grand canonical ensemble, the pair correlation function is a sum over statistically weighted pair correlation functions of each of the states in the ensemble labeled with collective index, $\nu$ (in this section, we follow notation and convention of Refs.~\cite{P85b} and~\cite{PK98b}). 
 A state $\Psi_{\lambda,\nu}$ has particle number $N_\nu$, energy $E_{\nu}$, and corresponds to $\lambda$-scaled interaction. If its weight in the ensemble is denoted as
\begin{equation}
w_{\lambda,\nu}=\frac{e^{-\beta(E_{\lambda,\nu}-\mu N_\nu)}}{\sum_\nu e^{-\beta(E_{\lambda,\nu}-\mu N_\nu)}},
\end{equation}
the ensemble average of the exchange-correlation hole density is
\begin{equation}
\left\langle n_{xc}^\lambda({\bf r},{\bf r'})\right\rangle=\sum_\nu w_{\lambda,\nu} n_{xc,\nu}^\lambda({\bf r},{\bf r'}).
\end{equation}

However, the exchange-correlation hole function used to obtain ${\cal F}^{\tau}_{xc}$ through a $\lambda$ integration requires the addition of
more complicated terms~\cite{PK98b}:
\begin{align}
\label{truehole}
n_{xc}^\lambda({\bf r},{\bf r'})&= \left\langle n_{xc}^\lambda({\bf r},{\bf r'})\right\rangle\notag\\ 
&+\sum_\nu w_{\lambda,\nu}\frac{[n_{\lambda,\nu}({\bf r})-n({\bf r})]}{n({\bf r})}[n_{\lambda,\nu}({\bf r'})+n_{xc,\nu}^\lambda({\bf r},{\bf r'})],
\end{align}
where $n_{xc,\nu}^\lambda$ is the usual exchange-correlation hole corresponding to $\Psi_{\lambda,\nu}$ with particle density $n_{\lambda,\nu}$.

Thus, the sum rule stated in the ground state gets modified as follows~\cite{Pb85}
\begin{equation}
\int d^3 r'~n_{xc}^\lambda({\bf r},{\bf r'})=-1+\sum_\nu w_{\lambda,\nu} \frac{n_{\lambda,\nu}({\bf r})}{n({\bf r})} [N_\nu - \langle N\rangle].
\end{equation}
The last expression shows that the sum rule for the thermal exchange-correlation hole accounts for an additional term due to particle number fluctuations.  
Worse still, this term carries along with it state-dependent, and therefore system-dependent, quantities.
This is an important warning that standard methodologies for producing reliable ground-state functional approximations must be 
properly revised for use in the thermal context.

\sec{Conclusion}
\label{sec:7}

Thermal density functional theory is an area ripe for development in both fundamental theory and the construction of approximations because of rapidly expanding applications in many areas. 
Projects underway in the scientific community include construction of temperature-dependent GGAs~\cite{PPB13}, exact exchange methods for non-zero temperatures~\cite{GCG10}, orbital-free approaches at non-zero temperatures~\cite{KST12}, and continued examination of the exact conditions that may guide both of these developments~\cite{PPB13}.  In the world of warm dense matter, simulations 
are being performed, often very successfully~\cite{PSTD12}, generating new insights into both materials science and the quality of our current approximations~\cite{PPB13b}.  
As discussed above, techniques honed for zero-temperature systems should be carefully considered before being applied to thermal problems.  Studying exact properties of functionals may guide efficient progress in application to warm dense matter.  In context, thermal DFT emerges as as a clear and solid framework that provides users and developers practical and formal tools of general fundamental relevance.

\sec{Acknowledgments}
We would like to thank the Institute for Pure and Applied Mathematics for organization of Workshop IV: Computational Challenges in Warm Dense Matter and for hosting APJ during the Computational Methods in High Energy Density Physics long program. APJ thanks the U.S. Department of Energy (DE-FG02-97ER25308), SP and KB thank the National Science Foundation (CHE-1112442), and SP and EKUG thank European Community's FP7, CRONOS project, Grant Agreement No. 280879.

\pagestyle{plain}
\bibliography{tDFTArxMaster}

\end{document}